\documentstyle[12pt]{article}

\catcode`\^^?=9 % ignore

\def\bb{\bigskip\bigskip}
\def\bbb{\bigskip\bigskip\bigskip}
\def\s{\smallskip}
\def\ce{\centerline}
\def\ve{\vfill\eject}

\def\be{\begin{equation}}
\def\ee{\end{equation}}
\def\ba{\begin{eqnarray}}
\def\ea{\end{eqnarray}}
\def\ban{\begin{eqnarray*}}
\def\ean{\end{eqnarray*}}

\begin{document}
\rightline {UG - 9/95}
\bb
\ce {\bf DERIVATIVE EXPANSION FOR THE ONE-LOOP EFFECTIVE}
\ce {\bf LAGRANGIAN IN QED}

\bb
\ce{\it  V.P. Gusynin$^{1,2}$ and I.A.Shovkovy$^{1,3}$}
\s
\ce{\it $^1$ Bogolyubov Institute for Theoretical Physics}
\ce{\it 252143, Kiev, Ukraine}
\s
\ce{\it $^2$ Institute for Theoretical Physics}
\ce{\it University of Groningen, 9747 AG Groningen, The Netherlands}
\s
\ce{\it $^3$ Department of Applied Mathematics}
\s
\ce{\it University of Western Ontario, London, Canada N6A 5B7}
\bbb
\ce {\bf Abstract}
\s
The derivative expansion of the one-loop effective Lagrangian in QED$_4$ is
considered. The first term in such an expansion is the famous Schwinger
result for a constant electromagnetic field. In this paper we give an
explicit expression for the next term containing two derivatives of the
field strength $F_{\mu\nu}$. The results are presented for both fermion and
scalar electrodynamics. Some possible applications of an inhomogeneous external
field are pointed out.
\ve
\section{\bf Introduction}

The effective Lagrangian for a constant electromagnetic field was
calculated by Schwinger more than forty years ago \cite{Sch}. The Schwinger
Lagrangian has the virtue of being nonpertubative, taking into account the
effects of the background field to all orders. However, its validity is
restricted to constant fields and, since that time, much efforts have been
devoted to finding a systematic way to generalize the result for strong slowly
varying fields \cite{Lee,DHok} or for the fields localized in a finite
region of space \cite{Martin}. Such a generalization would be of great
interest from the physical point of view, since it can be applied to studying
instabilities (such as spontaneous symmetry breakdown) in quantum field theory
in external field \cite{GMS1,GMS2}, or it can be used for the explanation of
the narrow $e^-e^+$ peaks observed in heavy--ion scattering experiments
\cite{Caldi}, where strong varying electromagnetic fields are indispensable
ingredients. Also, nonperturbative information on the fermion determinant,
closely connected with the one--loop effective action, is very important
because fermion determinants produce an effective measure for gauge fields
when the fermions are integrated out \cite{Fry}.

In the present paper, we calculate the low-energy effective action for
fermions and bosons in QED$_4$ as a derivative expansion around
nonvanishing field strength
\begin{equation}
{\cal L}_{eff}={\cal L}_{0}(F_{\mu\nu})+\partial_{\lambda}F_{\alpha\beta}
\partial_{\gamma}F_{\sigma\delta}{\cal L}_{1}^{\lambda\alpha\beta\gamma
\sigma\delta}(F_{\mu\nu})+\ldots,
\label{eq:derivat}
\end{equation}
where ${\cal L}_{k}$'s are some local functions of the field--strength tensor
$F_{\mu\nu}$. In the derivative expansion (\ref{eq:derivat}) the first
term ${\cal L}_{0}$ is the famous Schwinger result and we compute the
next term ${\cal L}_{1}$.

Recently, an analogous calculation has been performed for QED in $2+1$
dimensions \cite{DHok}. As for $3+1$ dimensions, we mention the paper
\cite{Lee} where the heat--kernel method was adopted to obtain the derivative
expansion (\ref{eq:derivat}). However, it is impossible to compare our
results with those in \cite{Lee} because some important functions in
the cited paper were not given explicitly. We comment more on comparison
with results of Refs.\cite{Lee,DHok} at the end of the paper.

\bb

\section{\bf Derivative Expansion of the One--Loop Effective Action in
\boldmath\mbox{$\rm QED_4$}}

The  one--loop effective action in QED reduces to computing the fermion
determinant
\begin{eqnarray}
iW^{(1)}(A)&=&\ln {\rm Det}(i\hat{{\cal D}}-m)=\frac{1}{2}\ln
{\rm Det}\left(-{\cal D}^2_{\mu}
+\frac{e}{2}\sigma_{\mu\nu}F^{\mu\nu}+m^2\right)=\nonumber\\
&=&\frac{1}{2}\int d^4x\langle x|tr\ln\left(-{\cal D}^2_{\mu}+
\frac{e}{2}\sigma_{\mu\nu}F^{\mu\nu}+m^2\right) |x\rangle.
\label{eq:ActGen2}
\end{eqnarray}
Here $\hat{{\cal D}}=\gamma^{\mu}{\cal D}_{\mu}$ and ${\cal D}_{\mu}=
\partial_{\mu}+ieA_{\mu}$ is the covariant derivative, $\sigma_{\mu\nu}=
i/2[\gamma_{\mu},\gamma_{\nu}]$ and $tr$ refers to spinor indices of Dirac
matrices $\gamma_{\mu}$; $|x\rangle$ are eigenstates for some
self--conjugated operators $x_{\mu}$ of coordinates. Throughout the paper
we use the Minkowski metric $\eta_{\mu\nu}=(1,-1,-1,-1)$.

With use of the identity
$\ln A=-\int_0^{\infty}\exp(-i\tau A)d\tau /\tau$ for introducing the
proper--time coordinate, Eq.(\ref{eq:ActGen2}) can be represented through the
diagonal matrix elements of the operator $U(\tau)=\exp(-i\tau H)$:
\begin{equation}
W^{(1)}(A)=\frac{i}{2}\int d^4x \int\limits^{\infty }_{0}\frac{d\tau}{\tau}
e^{-im^2\tau}tr\langle x|\exp(-i\tau H)|x\rangle,
\label{eq:ActGen3}
\end{equation}
where the second order differential operator $H$ is given by
\begin{equation}
H=\Pi_{\mu}\Pi^{\mu}+\frac{e}{2}\sigma_{\mu\nu}F^{\mu\nu}(x),\qquad
\Pi_{\mu}=-i{\cal D}_{\mu}.
\label{eq:operH}
\end{equation}
The evolution operator $U(\tau)$ satisfies the Schr\"{o}dinger equation
\begin{eqnarray}
i\frac{dU(\tau)}{d\tau}=HU(\tau)
\end{eqnarray}
together with the initial condition $U(0)=1$. In Euclidean space this
operator is closely connected with the heat kernel operator.  Because the
matrix elements of $U(\tau)$ cannot be calculated for arbitrary external
fields, several approximate regular schemes have been developed: the technique
of asymptotic expansion at small values of the proper time ($\tau\to 0$)
[9--13], covariant perturbative theory \cite{BaVi2,BGVZ} and, at last,
pseudodifferential operator technique [16--19]. The asymptotic expansion at
$\tau\to 0$ is equivalent to the asymptotic expansion of the effective action
in inverse powers of mass parameter $m^2$. The coefficients of this expansion
are the space--time integrals of local invariants, $E_n(x,x)$
($n=0,1,2,\ldots$), of growing powers in field strenth $F_{\mu\nu}$ and its
derivatives, called DeWitt--Seeley--Gilkey coefficients.  The local
(Schwinger--DeWitt) expansion is, therefore, an asymptotic approximation valid
for weak and slowly varying background fields. The covariant perturbation
theory of Barvinsky and Vilkovisky deals with the case of weak and rapidly
varying fields. It corresponds to summing up all terms in the Schwinger--DeWitt
series with a given power of field strength and any number of its derivatives.
The covariant perturbation theory like the small proper--time expansion, fails
in the case of large fields. An approximation scheme complimentary to the above
mentioned ones would correspond to strong and slowly varying fields (derivative
expansion).  The lowest--order approximation of such a scheme is the case of
covariantly constant background fields. In principle, the most general approach
of pseudodifferential operators permits one to develop such a derivative
expansion (see, for example, \cite{GuKu} and \cite{Lee}), but in the present
paper we shall follow another approach to derive the expansion
(\ref{eq:derivat}).

The matrix elements $\langle x|\exp(-i\tau H)|x\rangle$ entering the
right--hand side of Eq.(\ref{eq:ActGen3}) may be treated as the
matrix elements of
the evolution operator of a spinning particle, $\tau$ and $H$ being the
proper time and Hamiltonian of the particle. The corresponding canonical
momenta are $P_{\mu}$'s which obey the commutation relations
$[x_{\mu},P^{\nu}]=i\delta^{\nu}_{\mu}$ and
$\langle x|P_{\mu}|y\rangle=-i\partial_{\mu}\delta(x-y)$.
Following the standard method
\cite{Frad}, the transition amplitude $\langle z|U(\tau)|y\rangle$ between
points $x(0)=y$ and $x(\tau)=z$ can be written in terms of a path integral over
c-number ($x_{\mu}(t)$) and Grassmann ($\psi_{\mu}(t)$) coordinates:
\begin{equation}
tr\langle z|U(\tau)|y\rangle=N^{-1}\int {\cal D}[x(t),\psi (t)]\exp\left\{i\int
\limits^{\tau}_{0} dt\left[L_{bos}(x(t))+L_{fer}(\psi(t),x(t))\right]\right\},
\label{eq:evol}
\end{equation}
where $N$ is a normalization factor, and
\begin{equation}
L_{bos}(x)= -\frac{1}{4}\frac{dx_{\nu }}{dt}\frac{dx^{\nu }}{dt}-eA_{\nu }(x)
\frac{dx^{\nu }}{dt},
\label{eq:L_bos}
\end{equation}
\begin{equation}
L_{fer}(\psi ,x)= \frac{i}{2} \psi _{\nu }\frac{d\psi ^{\nu }}{dt}-
ie \psi ^{\nu }\psi ^{\lambda } F_{\nu \lambda }(x).
\label{eq:L_fer}
\end{equation}
The integration in Eq.(\ref{eq:evol}) goes over trajectories $x^{\mu}(t)$ and
$\psi ^{\mu }(t)$ parametrized by $t\in [0,\tau]$ and obeying the boudary
conditions
\begin{equation}
x(0)=y,\qquad x(\tau)=z,\qquad \psi (0)=-\psi (\tau).
\end{equation}
The Lagrangian $L_{bos}(x)+L_{fer}(\psi,x)$ may be treated as Lagrangian of
the relativistic spinning particle in external electromagnetic field
\cite{Frad,Fain,Holt}.\footnote{For a more sophisticated representation in
terms of path-integral with reparametrization-invariant Lagrangian, see
\cite{Frad,Holt}. We note that from the point of view of
further using the first quantized path integral in applications it would be
quite desirable to have a representation for the evolution operator in
terms of boson operators only. Since $\gamma$-matrices enter the
operator (\ref{eq:operH}) through $\sigma_{\mu\nu}$, satisfying $SO(3,1)$
Lie algebra, such a representation should exist.}
 The first--quantized representation for the effective action,
given by Eqs.(\ref{eq:ActGen3}), (\ref{eq:evol})--(\ref{eq:L_fer}), can be used
as a starting point for developing approximate schemes for computing the
expansion in the case of arbitrary external fields. The path integral
(\ref{eq:evol}) is the particle theory analogue of the Polyakov path
integral in string theory, and motivated by this fact, Strassler
\cite{Strass} has developed the first--quantized
perturbation theory for calculating the one--loop effective action. Moreover,
such an approach to ordinary field theory, based on path integrals defined on
one--dimensional world lines, can be extended to the evaluation of Feynman
diagrams for Green's functions at arbitrary loop order \cite{McKeon} and to the
computation of form--factors \cite{Schm}. In the last case it was shown how
the higher derivative terms appearing in the effective action can be
effectively  summed up and lead to the results obtained in the heat kernel
approach of Barvinsky and Vilkovisky \cite{BaVi2}.

In order to generate the expansion (\ref{eq:derivat}) from the path integral
(\ref{eq:ActGen3}) it is
convenient to choose a special gauge condition for the potential $A_{\mu}(x)$,
namely the Fock--Schwinger gauge \cite{Fock}
\begin{equation}
(x^{\nu }-y^{\nu })A_{\nu }(x)=0,
\end{equation}
which leads to the series
\begin{eqnarray}
A_{\nu }(x)&=&\frac{1}{2}(x^{\lambda }-y^{\lambda })F_{\lambda \nu }(y)+
\frac{1}{3}(x^{\lambda }-y^{\lambda })(x^{\sigma }-y^{\sigma })
\partial _{\sigma }F_{\lambda \nu }(y)\nonumber\\
&+&\frac{1}{8}(x^{\lambda }-y^{\lambda })(x^{\sigma }-y^{\sigma })
(x^{\mu }-y^{\mu })
\partial _{\sigma }\partial _{\mu }F_{\lambda \nu }(y)+\ldots\nonumber\\
&=&\sum^{\infty}_{n=0}\frac{(x^{\lambda }-y^{\lambda })(x^{\nu _{1}}-
y^{\nu _{1}})\ldots
(x^{\nu _{n}}-y^{\nu _{n}})}{n!(n+2)}\partial _{\nu _{1}}\partial _{\nu_2}
\ldots\partial _{\nu _{n}}F_{\lambda \nu }(y).
\label{eq:Anu}
\end{eqnarray}
Carring  out  the  change  of the variable $x(t)$  for
$x'(t)=x(t)-y$ in the path integral (\ref{eq:evol}) (henceforth we omit
the prime) and substituting  (\ref{eq:Anu})  into  (\ref{eq:evol}), we  obtain
\begin{eqnarray}
&&tr\langle z|U(\tau)|y\rangle=N^{-1}\int D[x(t),\psi (t)]
\exp\Bigg[i\int\limits^\tau_0 dt
\Bigg(-\frac{1}{4}\frac{d x_\nu}{dt}\frac{d x^\nu}{dt}
-\frac{e}{2}x^{\lambda }F_{\lambda \nu }(y)\nonumber\\
&&\times\frac{d x^\nu}{dt}+L_{2}(x)\Bigg)\Bigg]
\exp\Bigg[i\int\limits^{\tau}_{0}dt\Bigg(\frac{i}{2}\psi_{\nu}
\frac{d\psi^\nu}{dt}-ie\psi^{\nu }\psi^{\lambda }F_{\nu \lambda }(y)
+L_{3}(x,\psi )\Bigg)\Bigg].
\label{eq:trU1}
\end{eqnarray}
The new  boundary  conditions for $x(t)$  are $x(0)=0$ and $x(\tau)=z-y$.
In Eq.(\ref{eq:trU1}) $F_{\mu \nu }$  does  not  depend  on $x(t)$.
The expressions for the interacting terms, $L_{2}(x)$   and $L_{3}(x,\psi )$,
containing  derivatives  of $F_{\mu \nu }$  with  respect  to  coordinates
take the form:
\begin{eqnarray}
L_{2}(x)&=& \sum^{\infty}_{n=1}\frac{eF_{\nu _{0}\nu _{1},\nu _{2}\ldots
\nu _{n+1}}}{n! (n + 2)}\frac{d x^{\nu_0}}{dt}x^{\nu _{1}}
(t)\ldots x^{\nu_{n+1}}(t) \nonumber\\
&=&\frac{e}{3}F_{\nu \lambda ,\sigma }\frac{d x^\nu}{dt}
x^{\lambda } x^{\sigma }+\frac{e}{8}F_{\nu \lambda ,\sigma \kappa }
\frac{d x^\nu}{dt} x^{\lambda } x^{\sigma } x^{\kappa }
+ \ldots,\\
L_{3}(x,\psi )&=&-\sum^{\infty}_{n=1}\frac{i}{n!}eF_{\lambda \mu ,
\nu _{1}\ldots
\nu _{n}} \psi ^{\lambda }(t)\psi  ^{\mu }(t)x^{\nu _{1}}(t)\ldots
x^{\nu_{n}}(t)\nonumber\\
&=&-ieF_{\nu \lambda ,\sigma } \psi ^{\nu }\psi ^{\lambda }x^{\sigma }-
\frac{ie}{2}F_{\nu \lambda ,\sigma \kappa } \psi ^{\nu }\psi ^{\lambda }
x^{\sigma }x^{\kappa }+ \ldots.
\end{eqnarray}
Here we use a conventional notation for derivatives:
$F_{\lambda \mu ,\nu _{1}\nu_{2}\ldots\nu _{n}}(x) =\\ \partial _{\nu _{1}}
\partial _{\nu_2}\ldots\partial _{\nu _{n}}F_{\lambda \mu }(x)$.

Introducing  c--number  and  Grassmann  external   sources,  we
 rewrite  (\ref{eq:trU1})  as
\begin{eqnarray}
tr\langle z|U(\tau)|y\rangle&=&\exp \Bigg[i \int\limits^{\tau}_{0} dt
\Bigg(L_{2}\left(\frac{1}{i}\frac{\delta}{\delta \eta (t)}\right)
+L_{3}\left(\frac{1}{i}\frac{\delta}{\delta \eta (t)} ,-
\frac{\delta}{\delta \xi (t)} \right)\Bigg)\Bigg]\nonumber\\
&\times&\left.Z_{\tau}[\eta ,\xi ](z;y)\right|_{\eta =0,\xi =0},
\label{eq:trU2}
\end{eqnarray}
where the generating functional is defined by
\begin{eqnarray}
&&Z_{\tau}[\eta ,\xi ](z;y)=N^{-1}\int D[x(t),\psi (t)]\exp\Bigg[
\frac{i}{2}\int
\limits^\tau_0 dt\Bigg(-\frac{1}{2}\frac{d x_\nu}{dt}\frac{d x^\nu}{dt}-
ex^{\lambda }F_{\lambda \nu }(y)\nonumber\\
&&\times\frac{d x^\nu}{dt}+2\eta_{\nu}x^{\nu}\Bigg)\Bigg]
\exp\Bigg[-\frac{1}{2}\int\limits^{\tau}_{0}dt\Bigg(\psi_{\nu}
\frac{d\psi^\nu}{dt}-2e\psi^{\nu }\psi^{\lambda }F_{\nu \lambda }(y)
+2\xi_{\nu}\psi^{\nu}\Bigg)\Bigg].
\label{eq:Gfunct}
\end{eqnarray}
Since the expression in the exponent of the right--hand side of
(\ref{eq:Gfunct}) is quadratic in the variables $x$ and $\psi$, the
calculation of the generating functional $Z_{\tau}[\eta ,\xi ](z;y)$
can be reduced to the calculation of the determinants of the following
one--dimensional differential operators:
\begin{eqnarray}
O_1=\frac{\eta_{\mu\nu}}{2}\frac{d^2}{dt^2}-eF_{\mu\nu}\frac{d}{dt}\qquad
\mbox{and}\qquad O_2=i\eta_{\mu\nu}\frac{d}{dt}-2ieF_{\mu\nu},
\end{eqnarray}
defined on the interval $[0,\tau]$ with the periodic and antiperiodic
boudary conditions, respectively. Taking into account the initial condition
\begin{equation}
Z_{\tau=0}[\eta ,\xi ](z;y)=\delta (z-y),
\end{equation}
which is equivalent to the operator equality $U(0)=1$, and introducing two
independent invariants constructed from the field tensor $F_{\mu\nu}$:
\begin{eqnarray}
K_{+}=\sqrt{\sqrt{ {\cal F}^2+{\cal G}^2 }+{\cal F} }\qquad\mbox{and}\qquad
K_{-}=\sqrt{\sqrt{ {\cal F}^2+{\cal G}^2 }-{\cal F} },
\end{eqnarray}
with
\begin{eqnarray}
{\cal F}=-\frac{1}{4} F^{\mu \nu }F_{\mu \nu }&\mbox{and}&{\cal G}=\frac{1}{4}
\epsilon ^{\mu \nu \lambda \kappa } F_{\lambda \kappa } F_{\mu \nu } ,
\end{eqnarray}
one can represent the result of the integration in (\ref{eq:Gfunct}) at
coincident arguments $z=y=x$ as
\begin{eqnarray}
Z_{\tau}[\eta ,\xi ](x;x)&=&-\frac{i}{4\pi ^{2}\tau^{2}}(e\tau
K_{+}) (e\tau K_{-})\coth(e\tau K_{+})\cot(e\tau K_{-})\nonumber\\
&\cdot&\exp\left(\frac{i}{2}S^{bos}_{cl}[\eta ]-\frac{1}{2}S^{fer}_{cl}
[\xi ]\right) ,
\label{eq:Gfun}
\end{eqnarray}
where the quadratic in external sourses forms, $S^{bos}_{cl}$ and
$S^{fer}_{cl}$, are defined by
\begin{eqnarray} S^{bos}_{cl}[\eta
]&=&\int\limits^{\tau}_{0}dt_{1} \int\limits^{\tau}_{0} dt_{2} \eta _{\nu
}(t_{1}) D^{\nu \lambda }(t_{1},t_{2}) \eta _{\lambda } (t_{2}) ,\\
S^{fer}_{cl}[\xi ]&=&\int\limits^{\tau}_{0}dt_{1}\int\limits^{\tau}_{0}
dt_{2} \xi _{\nu }(t_{1}) S^{\nu \lambda }(t_{1},t_{2}) \xi _{\lambda }
(t_{2}) ,
\end{eqnarray}
The Green functions entering the last expression are defined by the formulae:
\begin{eqnarray}
D^{\nu \lambda }(t_{1},t_{2})&=& \sum^4_{j=1} A^{\nu \lambda }_{(j)}
\frac{1}{2ef_{j}} \Bigg(\epsilon (t_{1}-t_{2}) \left(1-\exp[2ef_{j}
(t_{1}-t_{2})]\right)\nonumber\\
&+&\coth(ef_{j}\tau)\left(1 + \exp[2ef_{j}(t_{1}-t_{2})]\right)\nonumber\\
&-&\sinh^{-1}(ef_{j}\tau)\left(\exp[ef_{j}(\tau-2t_{2})] +
\exp[ef_{j}(2t_{1}-\tau)]\right)\Bigg),
\label{eq:Dmunu}
\end{eqnarray}
\begin{equation}
S^{\nu \lambda }(t_{1},t_{2})= \sum^{4}_{j=1}  A^{\nu \lambda }_{(j)}
\frac{1}{2}\left(\epsilon (t_{1}-t_{2})-\tanh(ef_{j}\tau)\right)\exp[2ef_{j}
(t_{1}-t_{2})].
\label{eq:Smunu}
\end{equation}
Here we use the matrices $A^{\nu \lambda }_{(j)}$ which were originaly
introduced  in \cite{Batal}. The explicit expressions for them are
\begin{equation}
A_{\mu \nu (j)}=\frac{-\bar{f}^{2}_{j} g_{\mu
\nu } + f_{j} F_{\mu \nu } + F^{2}_{\mu \nu} - i \bar{f}_{j}
\stackrel{*}{F}_{\mu \nu} } { 2 ( f^{2}_{j} - \bar{f}^{2}_{j} )},
\label{eq:matrA}
\end{equation}
where
\begin{eqnarray*} f_{1}= iK_{-} ,\; f_{2}= - iK_{-} &,&
f_{3}= K_{+} ,\; f_{4}= - K_{+} ;\\ \bar{f}_{1}= - K_{+} ,\; \bar{f}_{2}=
K_{+} &,& \bar{f}_{3}= -i K_{-} ,\; \bar{f}_{4}= i K_{-}.
\end{eqnarray*}

The  main  property  of  matrices  (\ref{eq:matrA})  that  was   used   in
(\ref{eq:Dmunu}) and (\ref{eq:Smunu})  is
\begin{equation}
F^{\nu \lambda }A_{(i)\lambda \mu} = A^{\nu \kappa}_{(i)}
 F_{\kappa \mu} = f_{i}A^{\nu }_{(i)\mu }.
\end{equation}
Other useful properties of these matrices that will be used below are:
\begin{eqnarray}
\sum_{i=1}^{4}A^{\mu \nu }_{(i)}=\eta^{\mu \nu },
\qquad A_{\mu (i)}^{\;\mu }=1,
\qquad A_{(i)}^{\mu \nu }A_{\nu \lambda (j)}=\delta_{ij}
A^{\mu }_{\;\lambda (j)}.&&
\end{eqnarray}
As can be easily seen, in the case of vanishing field the propagators
$D^{\nu \lambda }(t_{1},t_{2})$ and $S(t_{1},t_{2})$ are reduced to the free
Green functions used in \cite{Schm}.

Substitution  of  (\ref{eq:Gfun})  and  (\ref{eq:Dmunu}), (\ref{eq:Smunu})
into (\ref{eq:trU2}) leads to   the expression  for $tr\langle x|U|x\rangle$.
The last step  is to expand  the exponent   in   terms   of   derivatives
and to   keep only two--derivative  terms. Therefore, one obtains:
\begin{eqnarray}
tr\langle x|U(\tau)|x\rangle&&=\Bigg(1 + i \int\limits^{\tau}_{0} dt[{\bf
V}_{2}(t)+{\bf W}_{2}(t)] -\frac{1}{2} \int\limits^{\tau}_{0}
\int\limits^{\tau}_{0}dt_{1}dt_{2} [{\bf V}_{1}(t_{1}){\bf V}_{1}(t_{2})
\nonumber\\&&\hspace{-2cm}+ {\bf W}_{1}(t_{1}){\bf W}_{1}(t_{2})] -\left.
\int\limits^{\tau}_{0} \int\limits^{\tau}_{0}dt_{1}dt_{2} {\bf V}_{1}(t_{1})
{\bf W}_{1}(t_{2})\Bigg)Z_\tau[\eta ,\xi ](x,x)\right|_{\eta=0,\xi=0},
\label{eq:trU3}
\end{eqnarray}
where
\begin{eqnarray}
{\bf V}_{1}(t)&=&\frac{i}{3} eF_{\nu \lambda ,\mu } \lim_{t_{0}\to t}
\frac{d}{dt_0}\frac{\delta^3}{\delta \eta _{\nu }(t_{0})
\delta \eta _{\lambda }(t)\delta \eta _{\mu }(t)},\nonumber\\
{\bf V}_{2}(t)&=&\frac{1}{8} eF_{\nu \lambda ,\mu \kappa} \lim_{t_{0}\to t}
\frac{d}{dt_0}\frac{\delta^4}{\delta \eta _{\nu }(t_{0})
\delta \eta _{\lambda }(t)\delta \eta _{\mu }(t) \delta \eta_{\kappa}(t)},
\nonumber\\
{\bf W}_{1}(t)&=&-eF_{\nu \lambda ,\mu }\frac{\delta ^{2}}{\delta \xi _{\nu }
(t)\delta \xi _{\lambda }(t)}\frac{\delta}{\delta \eta _{\mu }(t)},\nonumber\\
{\bf W}_{2}(t)&=&\frac{i}{2}eF_{\nu \lambda ,\mu \kappa}\frac{\delta ^{2}}
{\delta \xi _{\nu }(t)\delta \xi _{\lambda }(t)}
\frac{\delta^2}{\delta \eta _{\mu }(t)\delta \eta _{\kappa }(t)}.
\end{eqnarray}

Substituting the generating function (\ref{eq:Gfun}) which depends on the
Green functions (\ref{eq:Dmunu}) and (\ref{eq:Smunu}), one can rewrite
(\ref{eq:trU3}) in the form:
\begin{eqnarray}
&&tr\langle
x|U(\tau)|x\rangle=-\frac{i(e\tau K_{-}) (e\tau K_{+})\cot(e\tau
K_{-})\coth(e\tau K_{+})}{4\pi ^{2}\tau^{2}}\Bigg[1 -\frac{i}{8} eF_{\nu
\lambda,\mu \kappa}\nonumber\\
&&\hspace{-1cm}\times\int\limits_0^{\tau}dt \bigg(\dot{D}^{\nu\lambda
}(t,t)D^{\mu \kappa }(t,t)+\dot{D}^{\nu \mu }(t,t) D^{\lambda
\kappa }(t,t)+ \dot{D}^{\nu \kappa }(t,t)D^{\lambda \mu }(t,t) +4S^{\nu
\lambda }(t,t)\nonumber\\
&&\hspace{-1cm}\times D^{\mu \kappa }(t,t)\bigg)
-\frac{i}{18}e^{2}F_{\nu \lambda ,\mu } F_{\sigma \kappa ,\rho}
\int\limits_0^{\tau}\int\limits_0^{\tau}dt_1 dt_2 \bigg(9D^{\mu \rho }(1,2)
\left(S^{\kappa \sigma }(2,2)S^{\lambda \nu }(1,1)\right.\nonumber\\
&&\hspace{-1cm}\left.-2S^{\kappa \lambda }(2,1)S^{\sigma \nu }(2,1)\right)
+6S^{\sigma \kappa }(2,2)\left(\dot{D}^{\nu \lambda }(1,1)D^{\mu \rho }(1,2)
+\dot{D}^{\nu \mu }(1,1)D^{\lambda \rho }(1,2)\right.\nonumber\\
&&\hspace{-1cm}\left.+\dot{D}^{\nu \rho }(1,2)D^{\lambda \mu }(1,1)\right)
+\dot{D}^{\nu \lambda }(1,1)\dot{D}^{\sigma \kappa }(2,2)D^{\mu \rho }(1,2)
+2\dot{D}^{\nu \lambda }(1,1)\left(\dot{D}^{\sigma \rho }(2,2)\right.
\nonumber\\
&&\hspace{-1cm}\times\left.
D^{\mu \kappa  }(1,2)+\dot{D}^{\sigma \mu }(2,1)D^{\kappa \rho}(2,2)\right)
+\dot{D}^{\nu \mu }(1,1)\dot{D}^{\sigma \rho }(2,2)D^{\lambda \kappa }(1,2)
+2\dot{D}^{\nu \kappa }(1,2)\nonumber\\
&&\hspace{-1cm}\times\left(\dot{D}^{\sigma \rho }(2,2)
D^{\lambda \mu }(1,1)+\dot{D}^{\sigma \mu}(2,1)D^{\lambda \rho }(1,2)\right)
+\dot{D}^{\nu \kappa }(1,2)\dot{D}^{\sigma \lambda }(2,1)D^{\mu\rho}(1,2)
\nonumber\\
&&\hspace{-1cm}+\dot{D}^{\nu \rho }(1,2)
\dot{D}^{\sigma \mu }(2,1)D^{\lambda \kappa }(1,2)
+\ddot{D}^{\nu \sigma }(1,2)\left(D^{\lambda \mu }(1,1)D^{\kappa \rho }
(2,2)+D^{\lambda \kappa }(1,2)\right.\nonumber\\
&&\hspace{-1cm}\left.\times D^{\mu \rho}(1,2)+D^{\lambda \rho }(1,2)
D^{\mu \kappa }(1,2)\right)\bigg)\Bigg].
\end{eqnarray}
Here the dotted functions are defined by
\begin{eqnarray}
\dot{D}^{\mu \nu}(1,2)&\stackrel{def}{=}&\frac{\partial}{\partial t_1}
D^{\mu \nu}(t_1,t_2),\\
\ddot{D}^{\mu \nu}(1,2)&\stackrel{def}{=}&\frac{\partial ^2}
{\partial t_1\partial t_2}D^{\mu \nu}(t_1,t_2),\\
\dot{D}^{\mu \nu}(t,t)&\stackrel{def}{=}&\lim_{t_0\to t}\frac{\partial}
{\partial t_0}D^{\mu \nu}(t_0,t).
\end{eqnarray}
The straightforward, though tedious, computation gives the result
\begin{eqnarray}
&&tr\langle x|U(\tau)|x\rangle=-\frac{i}{4\pi ^{2}\tau^{2}}(e\tau K_{-})
(e\tau K_{+})\cot(e\tau K_{-})\coth(e\tau K_{+})\nonumber\\
&&\hspace{-7mm}\times\Bigg[1 -\frac{i}{8}
eF_{\nu \lambda,\mu \kappa}\sum_{j,l}
\bigg(C^{V}(f_{j},f_{l})\left(A^{\nu \lambda }_{(j)}A^{\mu \kappa }_{(l)}+
2 A^{\nu \mu }_{(j)}A^{\lambda \kappa \ }_{(l)}\right)+2C^{W}(f_{j},f_{l})
A^{\lambda \nu }_{(j)}A^{\mu \kappa }_{(l)}\bigg)\nonumber\\
&&\hspace{-7mm}-\frac{i}{18}e^{2}F_{\nu \lambda ,\mu } F_{\sigma \kappa,
\rho}\sum_{j,l,k}\Bigg(9C^{WW}_{1}(f_{j},f_{l},f_{k})A^{\kappa \sigma }_{(j)}
A^{\lambda \nu }_{(l)}A^{\mu \rho }_{(k)}+9C^{WW}_{2}(f_{j},f_{l},f_{k})
A^{\kappa \lambda }_{(j)}A^{\sigma \nu }_{(l)}\nonumber\\
&&\hspace{-7mm}\times A^{\mu \rho }_{(k)}+6C^{VW}_{1}(f_{j},f_{l},f_{k})
A^{\sigma \kappa }_{(j)}
\left(A^{\nu \lambda }_{(l)}A^{\mu \rho }_{(k)}+ A^{\nu \mu }_{(l)}
A^{\lambda \rho }_{(k)}\right)+6C^{VW}_{2}(f_{j},f_{l},f_{k})
A^{\sigma \kappa }_{(j)}A^{\nu \rho }_{(l)}A^{\lambda \mu }_{(k)}\nonumber\\
&&\hspace{-7mm}-C^{VV}_{1}(f_{j},f_{l},f_{k})\left(A^{\nu \lambda }_{(j)}
A^{\kappa \sigma }_{(l)}A^{\mu \rho }_{(k)}
+ A^{\nu \mu }_{(j)}A^{\kappa \rho }_{(l)}A^{\lambda \sigma }_{(k)}
+2 A^{\nu \lambda }_{(j)}A^{\kappa \rho }_{(l)}A^{\mu \sigma }_{(k)}\right)
\nonumber\\
&&\hspace{-7mm}-C^{VV}_{2}(f_{j},f_{l},f_{k})\left(A^{\nu \sigma }_{(j)}
A^{\kappa \lambda }_{(l)}A^{\mu \rho }_{(k)}+
A^{\nu \rho }_{(j)}A^{\kappa \mu }_{(l)}A^{\lambda \sigma }_{(k)}+
2 A^{\nu \sigma }_{(j)}A^{\kappa \mu }_{(l)}A^{\lambda \rho }_{(k)}\right)
\nonumber\\
&&\hspace{-7mm}- 2C^{VV}_{3}(f_{j},f_{l},f_{k}) \left(A^{\nu \lambda}_{(j)}
A^{\kappa \mu }_{(l)}A^{\sigma \rho }_{(k)}
+ A^{\kappa \rho }_{(j)}A^{\nu \sigma }_{(l)}A^{\lambda \mu }_{(k)}\right)
-C^{VV}_{4}(f_{j},f_{l},f_{k}) A^{\nu \kappa }_{(j)}A^{\lambda \mu }_{(l)}
A^{\sigma \rho }_{(k)}\nonumber\\
&&\hspace{-7mm}-C^{VV}_{5}(f_{j},f_{l},f_{k}) A^{\nu \kappa }_{(j)}
\left(A^{\lambda \sigma }_{(l)}A^{\mu \rho}_{(k)}+A^{\lambda \rho }_{(l)}
A^{\mu \sigma}_{(k)}\right)\Bigg)\Bigg],
\label{eq:trUfer}
\end{eqnarray}
where the coefficients $C_{i}^{XY}(\alpha,\beta,\gamma)$ (where $X$,
$Y\in \{V,W\}$) are given in Appendix.

In principle, the expression (\ref{eq:trUfer}) is our final result. It
remains only to substitute the last expression into (\ref{eq:ActGen3}) and
to renormalize field and charge according to the rule
\begin{equation}
\left(e^{phys}\right)^{2}=\frac{e^{2}}{1 + {\bf C} e^{2}},\qquad
A^{phys}_{\mu }=\sqrt{1+{\bf C}e^2}A_{\mu} ,
\end{equation}
\begin{equation}
{\bf C}=\frac{e^{2}}{12\pi }\int\limits^{\infty }_{0}\frac{d\tau}{\tau}
\exp(-im^{2}\tau) .
\end{equation}
Note that the renormalization used merely means performing some
subtraction (precisely the same as in the origional Schwinger paper
\cite{Sch}) from the term containing no derivatives of field strength and
changing all bare quantities for physical ones.

As was mentioned at the begining of the paper, the derivative expansion in
QED$_4$ was also studied in \cite{Lee}. The result of that paper, however,
was presented in an explicit form only for the special case of the
electromagnetic field:
\begin{equation}
{\cal G}=0,\qquad F^{\mu \nu }(x)=\Phi (x){\bf F}^{\mu \nu },
\label{eq:special}
\end{equation}
where $\Phi (x)$ is a scalar function  and ${\bf F}^{\mu \nu }$ is a constant
matrix.  Under the particular choice of the field
configuration given by (\ref{eq:special}) the derivative part of the diagonal
matrix elements (\ref{eq:trUfer}) reads
\begin{equation}
tr\langle x|U(\tau)|x\rangle_{der}=\frac{1}{(4\pi )^{2}\tau}
\frac{\partial _{\mu }\Phi \partial ^{\mu }\Phi }{\Phi ^{2}}(3\omega ^{2} Y^{4}
-3\omega Y^{3}-4\omega^{2}Y^{2}+3\omega Y+\omega ^{2}) ,
\label{eq:trUF}
\end{equation}
where $\omega  = e\tau \sqrt{2{\cal F}}$, $Y= \coth(\omega )$. We note that
this result was obtained in \cite{Lee} by means of symbolic calculations
on a computer using the REDUCE package.

Now turning to the calculation of the derivative expansion for the scalar
electrodynamics, one does not need to repeat all the calculations similar
to those made above. In order to see this, we recall that the effective
one--loop Lagrangian in this case reads
\begin{equation}
L_{bos}^{(1)}(x)=-i\int\limits^{\infty }_{0}\frac{d\tau}{\tau}
\langle x|U_{bos}(\tau)|x\rangle e^{-im^2\tau}.
\label{eq:Lbos}
\end{equation}
The evolution connected with the transition amplitude,
$\langle z|U_{bos}(\tau)|y\rangle$, is described now by the Hamiltonian
(compare with (\ref{eq:ActGen3}) and (\ref{eq:operH})):
\begin{equation}
H_{bos}=(i\partial _{\mu }-eA_{\mu}(x))(i\partial^\mu-eA^\mu(x)).
\end{equation}
Thus, omitting all terms originating from the fermion part in
(\ref{eq:evol}), {\em i.e.} putting $L_{3}=0$ in
(\ref{eq:trU1}), (\ref{eq:trU2}) and $S^{fer}_{cl}=0$ in (\ref{eq:Gfun}),
we come to the following expression
\begin{eqnarray}
&&\langle x|U_{bos}(\tau)|x\rangle=-\frac{i}{16\pi
^{2}\tau^{2}} \frac{(e\tau
K_{-})(e\tau K_{+})} {\sin(e\tau K_{-})\sinh(e\tau K_{+})}\nonumber\\
&&\times \Bigg[1-\frac{i}{8}eF_{\nu \lambda ,\mu \kappa }\sum_{j,l}
C^{V}(f_{j},f_{l})\bigg(A^{\nu \lambda }_{(j)}A^{\mu \kappa }_{(l)}
+ 2 A^{\nu  \mu }_{(j)}A^{\lambda \kappa \ }_{(l)}\bigg)
+\frac{i}{18}e^{2}F_{\nu \lambda ,\mu } F_{\sigma \kappa,\rho}\nonumber\\
&&\times\sum_{j,l,k}\Bigg(C^{VV}_{1}(f_{j},f_{l},f_{k})\bigg(A^{\nu
\lambda }_{(j)}A^{\kappa \sigma }_{(l)}A^{\mu \rho }_{(k)}
+ A^{\nu \mu }_{(j)}A^{\kappa \rho }_{(l)}A^{\lambda \sigma }_{(k)}
+2 A^{\nu \lambda }_{(j)}A^{\kappa \rho }_{(l)}A^{\mu \sigma
}_{(k)}\bigg)\nonumber\\
&&+C^{VV}_{2}(f_{j},f_{l},f_{k}) \bigg(A^{\nu \sigma }_{(j)}
A^{\kappa \lambda }_{(l)}A^{\mu \rho }_{(k)}
+A^{\nu \rho }_{(j)}A^{\kappa \mu }_{(l)}A^{\lambda \sigma }_{(k)}+
2 A^{\nu \sigma }_{(j)}A^{\kappa \mu }_{(l)}A^{\lambda \rho }_{(k)}\bigg)
\nonumber\\
&&+ 2C^{VV}_{3}(f_{j},f_{l},f_{k}) \bigg(A^{\nu \lambda }_{(j)}
A^{\kappa \mu }_{(l)}A^{\sigma \rho }_{(k)}+ A^{\kappa \rho }_{(j)}
A^{\nu \sigma }_{(l)}A^{\lambda \mu }_{(k)}\bigg)
+C^{VV}_{4}(f_{j},f_{l},f_{k})A^{\nu \kappa }_{(j)}A^{\lambda \mu }_{(l)}
A^{\sigma \rho }_{(k)}\nonumber\\
&&+C^{VV}_{5}(f_{j},f_{l},f_{k}) A^{\nu \kappa }_{(j)}
\bigg(A^{\lambda \sigma }_{(l)}A^{\mu \rho}_{(k)}+A^{\lambda \rho }_{(l)}
A^{\mu \sigma}_{(k)}\bigg)\Bigg)\Bigg].
\label{eq:trUbos}
\end{eqnarray}
The coefficients used here are the same as those that were used in
(\ref{eq:trUfer}). In the special case of the  electromagnetic field
of the form (\ref{eq:special}) we obtain
\begin{equation} \langle
x|U_{bos}(\tau)|x\rangle_{der}=\frac{1}{(8\pi )^{2}\tau}
\frac{\partial _{\mu }\Phi  \partial ^{\mu }\Phi }{\Phi ^{2}}
\left(\frac{3(\omega\coth\omega-1)}{\sinh^2\omega}+\omega\coth\omega-2\right)
\frac{\omega}{\sinh\omega}.
\end{equation}
To end this section, let us write down also the explicit expressions
for the effective Lagrangians in the case of a space--time--inhomogeneous
magnetic field, which is assumed to be small in comparison with the fermion
mass scale. As follows from our formulae (\ref{eq:ActGen3}),
(\ref{eq:trUfer}), (\ref{eq:Lbos}) and (\ref{eq:trUbos}), after performing
the integration over the proper--time parameter, the corresponding expressions
for two-derivative terms are:
\begin{eqnarray} {\cal
L}^{fer}_{der}&=&-\frac{e^2\partial_{\mu}B\partial^{\mu}B}{120\pi^2m^2}
\left(1-\frac{20}{21}\left(\frac{eB}{m^2}\right)^2+\ldots\right),\\
{\cal L}^{bos}_{der}&=&-\frac{e^2\partial_{\mu}B\partial^{\mu}B}{960\pi^2m^2}
\left(1-\frac{53}{21}\left(\frac{eB}{m^2}\right)^2+\ldots\right).
\end{eqnarray}
These results for four dimensional QED can be compared with those in the
case of three dimensional QED in the presence of a space--inhomogeneous
magnetic field.  For QED$_3$ (under the same conditions) it was found
\cite{DHok}
\begin{eqnarray} {\cal
L}^{fer}_{der}&=&\frac{\partial_{i}B\partial_{i}B}{(eB)^3}
\frac{m^3}{60\pi}\left(\frac{eB}{m^3}\right)^3+\ldots, \\
{\cal L}^{bos}_{der}&=&\frac{\partial_{i}B\partial_{i}B}{(eB)^3}
\frac{m^3}{240\pi}\left(\frac{eB}{m^3}\right)^3+\ldots. .
\end{eqnarray}
These are precisely the expressions that led the authors of the paper
\cite{DHok} to the conclusion that the appearance of inhomogeneous magnetic
fields is accompanied by a decrease of the vacuum energy. The latter would
imply a vacuum instability which tends to generate inhomogeneous magnetic
field in the vacuum state. However, the question of vacuum stability is
inaccessible in a derivative expansion, and in fact the infinite sum of
terms with derivatives drives a system towards a system with uniform
magnetic field \cite{DHok2}.

In the case of QED$_4$, a space--inhomogeneous magnetic field may lead
to decreasing the vacuum energy (recall that we use Minkowski metric) as well.
However, as in the three dimensional case, it may be only an artifact of
the derivative expansion approximation.

\bb
\section{\bf Conclusions}

The main result of our paper is the derivative expansion of the effective
action for both the fermion (see formulae (\ref{eq:ActGen3}) and
(\ref{eq:trUfer})) and the scalar (see formulae (\ref{eq:Lbos}) and
(\ref{eq:trUbos})) electodynamics in $3+1$ dimensions.  Note that
our results, in contrast to those obtained in the paper \cite{Lee}, are
derived in an explicit form for the most general electromagnetic field
configuration.

As for the application of the derivative expansion derived in the paper, it
could be useful in extending the dynamical  chiral symmetry breaking in
QED$_4$ \cite{GMS1,GMS2} by a magnetic field to the case where inhomogeneities
are present\footnote{In connection with this see the recent paper \cite{Par},
generalizing some results of Ref.\cite{GMS1,GMS2} to the case of an
inhomogeneous magnetic field.}. This would allow contact with the GSI
heavy--ion experiments \cite{Sal} where strong inhomogeneous electromagnetic
fields play an important role.

Another application could be connected with the existence of very strong,
inhomogeneous fields during the electroweak phase transition in the early
Universe \cite{Amb} which might essentially change the character of that
phase transition.

In conclusion we note that the technique used in the present paper can be
applied to generate the derivative expansion around covariantly constant
non--abelian and gravitational fields. The first term in such an expansion
has already been calculated by Savvidi \cite{Sav} with a result generalizing
Schwinger's old result \cite{Sch} to the non--abelian gauge--field. Recently,
such a calculation was extended to the case of nonvanishing, covariantly
constant, Riemann curvature by Avramidi \cite{Avr}.

\bb

\section{\bf Acknowledgements}
We wish to acknowledge D.G.C.McKeon from the University of Western Ontario
(London, Canada) for fruitful discussions. V.P.G. is grateful to the
members of the Institute for Theoretical Physics (the University of
Groningen, The Netherlands), especially D.Atkinson, for their hospitality
during his stay at Groningen University where the final version of the paper
was written.
V.P.G. would like to thank the Stiching FOM (Fundamenteel Onderzoek der
Materie), financially supported by the Nederlandse Organisatie voor
Wetenschappelijk Onderzoek, for its support.

The work was supported by the grant INTAS-93-2058 "East--West network in
constrained dynamical systems" and in part (for I.A.Sh.) by the
International Soros Science Education Program (ISSEP) through grant
PSU052143.

\bb

\section{\bf Appendix}

Here we give the functions that were used in (\ref{eq:trUfer}):
\begin{eqnarray*}
C^{W}(\alpha ,\beta ) &=& \tau^{2}\tanh(\alpha \tau)H(\beta \tau) ,\\
C^{V}(\alpha ,\beta ) &=& \alpha \tau^{3}H(\alpha \tau)H(\beta \tau)-
\frac{\alpha \tau}{\beta ^{2}-\alpha ^{2}}[H(\beta \tau)-H(\alpha \tau)],\\
\hspace{-3mm}C^{WW}_{1}(\alpha ,\beta ,\gamma ) &=&\frac{\tau^{3}}{8}
\tanh(\alpha \tau)\tanh(\beta \tau)H(\gamma \tau),\\
\hspace{-3mm}C^{WW}_{2}(\alpha ,\beta ,\gamma )
&=&\frac{\tau^{2}}{4}[\tanh(\alpha \tau)+
\tanh(\beta \tau)]\left(\frac{H(\alpha \tau+\beta \tau)-H(\gamma \tau)}
{\alpha +\beta -\gamma }-\frac{H(\gamma \tau)}{\alpha +\beta }\right),\\
\hspace{-3mm}C^{VW}_{1}(\alpha ,\beta ,\gamma ) &=& -\frac{\tau^{3}}{4}
\tanh(\alpha \tau)\left(\beta \tau H(\beta \tau)H(\gamma
\tau)-\frac{H(\beta \tau)-
H(\gamma \tau)}{\tau(\beta +\gamma )}\right) ,\\
\hspace{-3mm}C^{VW}_{2}(\alpha ,\beta ,\gamma ) &=& \frac{\tau^{2}\beta
\tanh(\alpha \tau)}
{2(\beta ^{2}-\gamma ^{2})} \left[H(\beta \tau)-H(\gamma \tau)\right] ,
\end{eqnarray*}
\begin{eqnarray*}
\hspace{-5mm}C^{VV}_{1}(\alpha ,\beta ,\gamma ) &=&\frac{\tau^{5}\alpha
\beta}{2}H(\alpha \tau)H(\beta \tau)H(\gamma \tau)
-\frac{\tau^{3}\alpha H(\alpha \tau)}{2(\beta -\gamma )} \left(H(\beta
\tau)-H(\gamma \tau)\right)\\
&-&\frac{\tau^{3}\beta H(\beta \tau)}{2(\alpha +\gamma )}
\left(H(\alpha \tau)-H(\gamma \tau)\right)-\frac{\tau}{2}
\frac{H(\alpha \tau)}{(\alpha +\gamma )(\alpha +\beta )} \\
&-&\frac{\tau}{2} \frac{H(\beta \tau)}{(\alpha +\beta )(\beta -\gamma )} +
\frac{\tau}{2} \frac{H(\gamma \tau)}{(\alpha +\gamma )(\beta -\gamma )} ,\\
\hspace{-5mm}C^{VV}_{2}(\alpha ,\beta ,\gamma ) &=& -\frac{\tau^{3}
\alpha \beta
H(\alpha \tau)H(\beta \tau)}{2(\alpha -\beta )(\alpha -\beta +\gamma )}+
\frac{\tau^{3}[2(\alpha -\beta )+\gamma \ ]}{2(\alpha -\beta )
(\alpha -\beta +\gamma) }\\
&\times&H(\gamma \tau)\left[\beta H(\beta \tau)-\alpha H(\alpha \tau)\right]\\
&+&\frac{\alpha \tau}{2}H(\alpha \tau)\left(\frac{2(\beta +\gamma )}
{(\alpha ^{2}-\beta ^{2})(\alpha ^{2}-\gamma ^{2})} - \frac{2\alpha -
\beta +\gamma}{(\alpha -\beta )^{2}(\alpha +\gamma )(\alpha -\beta +\gamma )}
\right)\\
&+&\frac{\beta \tau}{2}H(\beta \tau)\left(\frac{2(\gamma -\alpha )}
{(\alpha ^{2}-\beta ^{2})(\beta ^{2}-\gamma ^{2})}+ \frac{2\beta -\alpha -
\gamma}{(\alpha -\beta )^{2}(\beta -\gamma )(\alpha -\beta +\gamma )}\right)\\
&+&\frac{\tau}{2}H(\gamma \tau)\left(\frac{2(\alpha \beta +\gamma ^{2})}
{(\alpha ^{2}-\gamma ^{2})(\beta ^{2}-\gamma ^{2})} - \frac{\gamma}
{(\alpha +\gamma )(\beta -\gamma )(\alpha -\beta +\gamma )}\right)\\
&+&\frac{\tau}{2(\alpha -\beta )(\alpha -\beta +\gamma )},\\
\hspace{-5mm}C^{VV}_{3}(\alpha ,\beta ,\gamma ) &=& -\frac{\tau^{3}\alpha \beta
H(\alpha \tau)}{\beta ^{2}-\gamma ^{2}} \left[H(\beta \tau)-H(\gamma \tau)
\right]+\frac{\alpha \tau H(\alpha \tau)}{(\alpha -\beta )(\alpha ^{2}-
\gamma ^{2})}\\
&-&\frac{\tau}{\beta ^{2}-\gamma ^{2}} \left(\frac{\beta}{\alpha  -\beta }
H(\beta \tau)-\frac{\alpha \beta +\gamma ^{2}}{\alpha ^{2}-\gamma ^{2}}
H(\gamma \tau) \right) ,\\
\hspace{-5mm}C^{VV}_{4}(\alpha ,\beta ,\gamma ) &=& -\tau^{3} H(\beta\tau)
H(\gamma \tau)
 -\frac{2\tau\alpha ^{2}H(\alpha \tau)}{(\alpha ^{2}-\beta ^{2})(\alpha ^{2}-
 \gamma ^{2})}+\frac{\tau(\alpha ^{2}+\beta ^{2})H(\beta \tau)}
 {(\alpha ^{2}-\beta ^{2})(\beta ^{2}-\gamma ^{2})}\\
&+&\frac{\tau(\alpha ^{2}+\gamma ^{2})H(\beta \tau)}{(\alpha ^{2}-
\gamma ^{2})(\gamma ^{2}-\beta ^{2})},\\
\hspace{-5mm}C^{VV}_{5}(\alpha ,\beta ,\gamma ) &=&\frac{\tau^3 \alpha }{2}
\left(\frac{\alpha}{(\alpha +\beta )(\alpha +\beta +\gamma )}+\frac{1}
{\alpha +\gamma }\right) H(\alpha \tau)H(\beta \tau) \\
&+&\frac{\tau^3 \alpha }{2} \left(\frac{\alpha}{(\alpha +\gamma )
(\alpha +\beta +\gamma) }+\frac{1}{\alpha +\beta }\right) H(\alpha \tau)
H(\gamma \tau)\\
&+&\frac{\tau[H(\beta \tau)-H(\gamma \tau)]}{\beta ^{2}-\gamma^2}
+\frac{\tau^{3}}{2}\left(\frac{\beta \gamma (\beta +\gamma )}{\alpha +
\beta +\gamma }-2\alpha ^{2}\right) \frac{H(\beta \tau)H(\gamma \tau)}
{(\alpha +\beta )(\alpha +\gamma )}\\
&+&\frac{\alpha \tau H(\alpha \tau)
\left(2+\frac{\alpha +\beta}{\alpha +\gamma }+\frac{\alpha +\gamma}{\alpha +
\beta }\right) }{ 2(\alpha +\beta )(\alpha +\gamma )(\alpha +\beta +\gamma )}
+\frac{\tau}{2}\Bigg(\frac{H(\gamma \tau)+H(\beta \tau)}
{(\alpha +\beta )(\alpha +\gamma )}\\
&+& \frac{H(\gamma \tau)-H(\beta \tau)}
{(\alpha +\beta )(\beta -\gamma )}+ \frac{H(\gamma \tau)-H(\beta \tau)}
{(\alpha +\gamma )(\beta -\gamma )}
+\frac{\gamma H(\gamma \tau)}{(\alpha +\gamma )^{2}
(\alpha +\beta +\gamma )}\\
&&\hspace{-1cm}+\frac{\beta H(\beta \tau)}{(\alpha +\beta )^{2}
(\alpha +\beta +\gamma )}
- \frac{1}{(\alpha +\beta )(\alpha +\beta +\gamma) }-
\frac{1}{(\alpha +\gamma )(\alpha +\beta +\gamma )}
\Bigg).
\end{eqnarray*}
Here  we used the notation
\begin{eqnarray*}
H(x)&=&\frac{x\coth(x)-1}{x^{2}} .
\end{eqnarray*}
\ve

\end{document}